\documentclass[12pt]{article}
\usepackage{amsmath}
\usepackage{amsfonts}
\usepackage{amsthm, amssymb,multicol}
\usepackage{epsfig}
\font\tenscr=rsfs10 
\font\sevenscr=rsfs7 
\font\fivescr=rsfs5 
\skewchar\tenscr='177 \skewchar\sevenscr='177
\skewchar\fivescr='177
\newfam\scrfam \textfont\scrfam=\tenscr \scriptfont\scrfam=\sevenscr
\scriptscriptfont\scrfam=\fivescr

\def\centeron#1#2{{\setbox0=\hbox{#1}\setbox1=\hbox{#2}\ifdim
\wd1>\wd0\kern.5\wd1\kern-.5\wd0\fi
\copy0\kern-.5\wd0\kern-.5\wd1\copy1\ifdim\wd0>\wd1
\kern.5\wd0\kern-.5\wd1\fi}}
\def\ltap{\;\centeron{\raise.35ex\hbox{$<$}}{\lower.65ex\hbox{$\sim$}}\;}
\def\gtap{\;\centeron{\raise.35ex\hbox{$>$}}{\lower.65ex\hbox{$\sim$}}\;}

\title{Fourier Analysis of the BTZ Black Hole  \vskip .3in}
\author{\large{Ian M. Tolfree\,\thanks{\tt tolfree@pha.jhu.edu}}
\mbox{  }\ \ \\ \\
\emph{\small{Department of Physics and Astronomy}} \\
\emph{\small{Johns Hopkins University}} \\
\emph{\small{3400 North Charles St}}.\\
\emph{\small{Baltimore, MD 21218-2686}}}

\begin{document}
\baselineskip=17pt \pagestyle{plain}
\begin{titlepage}
\vskip -.4in \maketitle

\begin{abstract}
In this paper we extend our previous work regarding the role of the Fourier transformation in bulk to boundary mappings to include the BTZ black hole.  We follow standard procedures for modifying Fourier Transformations to accommodate quotient spaces and arrive at a bulk to boundary mapping in a black hole background.  We show that this mapping is consistent with known results and lends a new insight into the AdS/CFT duality.  We find that the micro-states corresponding to the entropy of a bulk scalar field are the Fourier coefficients on the boundary, which transform under the principal series representation of $SL(2,R)$.  Building upon this we present a toy model to analyze the implications of this for the origin of black hole entropy.  We find that the black hole micro-states live on the boundary and correspond to the possible emission modes of the black hole.

\end{abstract}
\thispagestyle{empty} \setcounter{page}{0}
\end{titlepage}

\section{Introduction}

	The BTZ black hole \cite{Banados:1992gq} is obtained by undertaking a coordinate transformation in AdS to the frame of an accelerated observer in order to generate the proper horizons, and then taking a quotient space of this coordinate system.  There exists previous work \cite{Steif:1993zv}, \cite{Lifschytz:1993eb} in which AdS Green's functions have been modified by the equivalence relation in order to find the BTZ Green's functions; both of these papers make use of the method of images.  They modify the Green's functions by inserting the equivalence relation defining the quotient space $\phi \rightarrow \phi + 2 \pi n$ into the AdS Green's functions and summing over the possible values of $n$.  In their Green's function they retain both the boundary coordinate $\phi'$ and the bulk coordinate $\phi$ over which the desired modification needs to be made.  

In the case of the real number line it is well known how to modify a Fourier transformation when the transformation over a quotient space is desired.  The simplest example is how to go from the Fourier transformation on a line to the transformation on a circle.  In doing so one does not modify the linear coordinate. Instead the continuous Fourier parameter, $k$, is replaced by a discrete one, $n$, that imposes the equivalence relation on the linear coordinate.  The substitution $k \longrightarrow \frac{n}{L}$ will make the points $x$ and $x + L$ equivalent.  The lesson to be learned from this is that it is the Fourier parameter that needs to be modified to impose the desired equivalence relation on the original space.  

One can show in the case of the line that such a modification is equivalent to the method of images by writing a function in terms of a Green's function propagating a source and Fourier transforming this.  The method of images can then be seen to project out integer values of $k$.  The advantage of modifying the transformation is that quantization of the Fourier parameter becomes manifest.  Because on AdS the boundary coordinates are the Fourier coordinates of the transformation, it is the boundary coordinates that need to be modified to exact the desired equivalence relation on the bulk coordinates. 
	
In \cite{Tolfree:2008am} we showed how the boundary CFT naturally arises as the quantized version of a gravitational theory in AdS by applying integral geometry techniques first developed by Gelfand, Graev, and Vilenkin.  The result can be simply stated that the boundary CFT is the momentum space of AdS as defined by the Fourier transformation.  This changes the interpretation of the duality because instead of dealing with boundary values of fields we are dealing with Fourier coefficients.  This became apparent when conditions the boundary fields are required to satisfy for normalization purposes arose as the natural properties of the Fourier coefficients.  This new interpretation of the duality in terms of Fourier transformations allows us to apply the existing machinery and intuition that already exists regarding Fourier transformations and the role they play in quantization to the duality.  

In this paper we use this approach to study BTZ black hole.  By suitably modifying not just the Green's functions but the entire transformation when going to the quotient space we show that the Fourier interpretation is consistent with known results (see \cite{Hamilton:2006fh}, \cite{Hamilton:2007wj}) and resolves the problem of divergences at the singularity posed in  \cite{Hamilton:2007wj}.  The modified Fourier transformation then tells us exactly which CFT states correspond to the quantum states of a scalar field around a BTZ black hole.   

We see that the micro-states contributing to the entropy of a scalar bulk field are the Fourier coefficients that live on the boundary.  Because of the way in which scalar fields interact with black holes, either by falling into them or stimulating emission from them in the case of superradiance, black holes must be able to create or annihilate states of a scalar field in the CFT-states represented by the Fourier coefficients.  We therefore propose that the micro-states contributing to black hole entropy are operators that can act on boundary states, such as a gauge background in the CFT, and that these are the states contributing to black hole entropy.  We also find in agreement with \cite{Guijosa:2003ze}, that the correspondence should be proposed in terms of principal series representations instead of highest weight representations \footnote{Although \cite{Guijosa:2003ze} refers to the dS/CFT correspondence this result is relevant because dS and euclidean AdS have the same group of isometries.}.  

The organization of the paper is as follows: first we review the Fourier Transformation on AdS, then we modify it to obtain the Fourier Transformation of the quotient space that is the BTZ black hole.  We then use the Green's functions (Fourier weights) to calculate the smearing functions that are used to map bulk states to boundary operators and analyze what this means for both bulk and boundary states.  Finally we speculate on the implication for black hole entropy and close with some concluding remarks.

\section{Review of the Fourier Transformation}
The Fourier transformation on $AdS_3$ with Euclidean signature is
\begin{equation}
f(x)=\frac{(-1)^{1}}{2^{2+1}(
\pi)^{1+\frac{1}{2}}\Gamma(1+\frac{1}{2})\textit{i}} \int_{a-\textit{i}\infty}^{a+\textit{i}\infty}
d\sigma  \times \frac{\Gamma(\sigma+2)}{\Gamma(\sigma)}
\int_{\textit{S}^\textit{2}}\Phi(\sigma,\xi)[x,\xi]^{-\sigma-2}d\xi
\end{equation}
where $\xi$ denote points on the boundary and $x$ denotes points on the hyperboloid.  $\sigma=\emph{i} \rho+a$ where $\rho$ is continuous and $a$ can be chosen such that the integral converges.  Here we take boundary to mean the points on the cone that are infinitely far from the hyperboloid in the embedding space.  $\xi$ lies on any contour intersecting all of the generators of the cone once.  For now we take it to lie on the intersection of the cone and an arbitrary plane such that the intersection is a circle of finite radius.  We can always take the radius to infinity to allow the contour to coincide what is usually meant by the boundary of AdS, but keeping the radius finite makes dealing with the representations a bit easier.

Since $\Phi(\sigma,\xi)$ is defined on a sphere, one can expand it in
terms of ultra-spherical harmonics.
\begin{equation}
\Phi(\sigma,\xi)=\sum_K a_K(\sigma)\Xi_K(\xi)
\end{equation}
where $K$ is a compact notation meaning all of the eigenvalues, e.g.
$K=(l,m_1,m_2...)$.  We can expand $\Phi(\sigma,\xi)$ in terms of spherical harmonics and
the Fourier transformation then reads:
\begin{equation}
f(x)= \frac{(-1)^{1}}{2^{2+1}(
\pi)^{1+\frac{1}{2}}\Gamma(1+\frac{1}{2})\textit{i}} \sum_K\int_{a-\textit{i}\infty}^{a+\textit{i}\infty}
d\sigma  \times \frac{\Gamma(\sigma+2)}{\Gamma(\sigma)}a_K(\sigma)
\int_{\textit{S}^\textit{2}} \Xi_K(\xi) [x,\xi]^{-\sigma-2}d\xi
\end{equation}
By changing the order of integration we can do the $\sigma$ integral.  The result is
\begin{equation}
f(x)= \frac{(-1)^{1}}{2^{2+1}(
\pi)^{1+\frac{1}{2}}\Gamma(1+\frac{1}{2})\textit{i}} 
  \times \frac{\Gamma(\Delta+2)}{\Gamma(\Delta)}
\sum_K \int_{\textit{S}^\textit{2}} a_K(\Delta)\Xi_K(\xi) [x,\xi]^{-\Delta-2}d\xi
\end{equation}
Where $\sum_K a_K(\Delta)\Xi_K(\xi)$ is what is usually referred to as $\phi_0$. 

Here it is helpful remind the reader of results from our previous work.  This looks like the construction of a bulk field in terms of boundary values of a field propagated into the bulk via the Green's function.  The insight the Fourier analysis adds is that one is not propagating the boundary value of a field into the bulk but is rather constructing the bulk field out of a weighted sum of Fourier coefficients that live only the boundary.  It is these coefficients that, when quantized, get promoted to operator status and create states in the boundary CFT.  In light of this we see that to properly obtain the CFT states corresponding to the BTZ black hole we need to modify the entire transformation.  The Green's function, which is really the Fourier weight, tells us how to modify the Fourier coordinates in the transformation (which \emph{are} the boundary coordinates) in order to obtain the transformation on the quotient space.

\section{Modification of the transformation}
To modify the transformation we start with the Fourier weight, which goes as  $[x, \xi]$, because this is the only spot that coordinates that lie on the hyperboloid appear in the transformation.  To do this we use the original BTZ coordinates in Lorentzian signature and work only in the region outside both event horizons.  The reader can check that any modification will be the same in all regions.  These coordinates are:
\begin{eqnarray}
u=\sqrt{A(r)}\cosh \tilde{\phi}(t, \phi) \\ \nonumber
x=\sqrt{A(r)}\sinh \tilde{\phi}(t, \phi)  \\ \nonumber
y=\sqrt{B(r)}\cosh \tilde{t}(t, \phi)  \\ \nonumber
v=\sqrt{B(r)}\sinh \tilde{t}(t, \phi) \\  
\end{eqnarray}
where \begin{eqnarray} A(r)=\emph{l}^2[\frac{r^2-r_-^2}{r_+^2-r_-^2}] \ \ B(r)=\emph{l}^2[\frac{r^2-r_+^2}{r_+^2-r_-^2}] \ \ \tilde{t}=\frac{1}{\emph{l}}(\frac{r_+t}{\emph{l}}-r_-\phi) \ \  \tilde{\phi}=\frac{1}{\emph{l}}(-\frac{r_-t}{\emph{l}}+r_+\phi). \end{eqnarray} and the equivalence relation defining the quotient space is $\phi \rightarrow \phi + 2 \pi n$.
We denote boundary coordinates by a $'$, and note they live on the cone.  Substituting these into the weight $[x, \xi]$ and simplifying yields the following 
\begin{eqnarray}
[x,\xi]=\emph{l}^2(\sqrt{\frac{(-r^2+r_-^2)(-{r'}^2+r_-^2)}{{(r_-^2-r_+^2)}^2}}\cosh(\frac{1}{\emph{l}}(\phi-\phi')r_+-(t-t')r_-)- \\ \nonumber\sqrt{\frac{(-r^2+r_+^2)(-{r'}^2+r_-^2)}{{(r_-^2-r_+^2)}^2}}\cosh(\frac{1}{\emph{l}}(\phi-\phi')r_--(t-t')r_+)).
\end{eqnarray}

In writing this down we note that $r'$ is really $r_b$, the section of the hyperboloid coinciding with the infinite radius contour on the boundary cone.  We take it to be this for ease, but keep in mind that any contour on the cone will do.  Because this location introduces infinities, we leave $r_b$ as $r'$ for the time being. 

The important thing to take away from this discussion is that we can now see how to modify the Fourier coefficient to impose the desired equivalence relation on $\phi$.  Recall one takes the quotient space to pass to the black hole background by making the $\phi$ coordinate periodic, $\phi \rightarrow \phi + 2 n \pi$.  The way to impose the equivalence relation is by modifying the boundary coordinate $\phi'$.  But because the Green's function goes as $\cosh$, which has period of $2 \pi \textit{i} n$, we must analytically continue $\phi$ and $\phi'$ before we can make the Green's function periodic in $\phi$.  To do so we need to impose restrictions on $\phi'$, which entails restricting it to $2 \pi n$.  Similarly one could view this as restricting the bulk coordinate to values of $2 \pi \textit{i} n$.  In either case the $\phi'$ integral needs to be replaced by a sum in the Fourier transformation, but first we must figure out how to write down the boundary coordinates in the transformation.

To see that that $r'=\infty$ lies on the cone we need to look at the definition of the BTZ surface in the embedding space.  The BTZ black hole is a particular parametrization of the hyperboloid $[x,x]=\emph{l}^2$. 
\begin{equation}
[x,x]=A(r)\cosh^2 \tilde{\phi}(t, \phi)-A(r)\sinh^2 \tilde{\phi}(t, \phi)+B(r)\cosh^2 \tilde{t}(t, \phi)-A(r)\sinh^2 \tilde{t}(t, \phi)=\emph{l}^2
\end{equation}
Factoring out either coefficient, $A(r)$ for example, yields the following:
\begin{equation}
[x,x]=\cosh^2 \tilde{\phi}(t, \phi)-\sinh^2 \tilde{\phi}(t, \phi)+\frac{B(r)}{A(r)}\cosh^2 \tilde{t}(t, \phi)-\frac{B(r)}{A(r)}\sinh^2 \tilde{t}(t, \phi)=\frac{\emph{l}^2}{A(r)}
\end{equation}
We can make use of the following two limits 
\begin{equation}
\lim_{r'\rightarrow\infty} \sqrt{\frac{B(r')}{A(r')}}=1 \; \; \;
\lim_{r'\rightarrow\infty} \sqrt{A(r')}=\infty
\end{equation} 
($B(r')$ could be used in the latter limit) to see that the quadratic surface becomes 
\begin{equation}
[x,x]=\cosh^2 \tilde{\phi}(t, \phi)-\sinh^2 \tilde{\phi}(t, \phi)-\cosh^2 \tilde{t}(t, \phi)+\sinh^2 \tilde{t}(t, \phi)=0
\end{equation}
which defines the conic surface $[\xi,\xi]=0$.  We can now drop the $r_b$ altogether and use the following coordinates to define the cone that is the Fourier space of the BTZ black hole :
\begin{eqnarray}
u'=\cosh \tilde{\phi}(t, \phi) \\ \nonumber
x'=\sinh \tilde{\phi}(t, \phi)  \\ \nonumber
y'=\cosh \tilde{t}(t, \phi)  \\ \nonumber
v'=\sinh \tilde{t}(t, \phi) .\\ 
\end{eqnarray}
It is these coordinates that we will use in the Fourier transformation for the boundary points.  The integration over the boundary now only needs to be taken over $t$ and $\phi$ and the limits on these coordinates are inherited from those on the hyperboloid.  

The convention in modifying the one dimensional Fourier transformation is to divide by the periodicity length, change the integral to a sum, and replace the continuous coefficient and differential by a discrete parameter.  Applying the same logic here, one finds the modified Fourier transformation describing a scalar field in a BTZ background is:   
\begin{equation} \label{grn}
f(x)=\frac{(-1)^{1}}{2^{2+1}(
\pi)^{1+\frac{1}{2}}\Gamma(1+\frac{1}{2})\textit{i}2 \pi^2}  \sum_{n=-\infty}^{n=\infty} \int_{a-\textit{i}\infty}^{a+\textit{i}\infty}
d\sigma  \times \frac{\Gamma(\sigma+2)}{\Gamma(\sigma)}  
\int_{\Gamma}\Phi(\sigma, t,2 \pi \emph{i} n))[x,\xi]^{-\sigma-2}dt
\end{equation}

The integral is only over the $t'$ variable now as taking the quotient space resulted in the quantization of $\phi'$.  To keep the notation neat we do not write out $x$ and $\xi$ explicitly in terms of coordinates although it is understood  when one wants to evaluate the integral they need to pick a coordinate system.  Writing the transformation like this allows us to use the same expansion in all three regions around the black hole: inside the inner horizon, between the two horizons, and outside the outer horizon.  One can check that in all three regions the modification of the transformation is identical.  The only difference in the expansion in the three regions is that one must plug in different expressions for the bulk coordinates to express the bulk function in each region.  The differences in the bulk coordinates in the different regions just entails the placements of minus signs in the radial portion of the coordinate expression. 

An interesting feature arising out of this formalism is that the boundary coordinates do not change for each region.  It seems that all three regions are mapped to the same support on the boundary.  This implies that although fields on either side of the horizon are causally disconnected in AdS, in the boundary theory there is nothing to stop them from talking to one another.  A subject of future work entails analyzing whether or not this feature allows information to leak out of the horizon via CFT states.  The probable origin for such a mechanism is that fields inside and outside the outer horizon are mapped to the same region on the boundary.  From there the fields originating inside the horizon could leak into the outer region from the boundary in two possible ways: either directly since it takes a finite amount of time to move from the boundary to the interior or through interaction with fields originating outside the horizon; a possible mode of interaction is the thermal bath the CFT is in.  In any case the door is opened for information to leak out of the horizon via the boundary theory.
	
Our last comment about the Fourier modification is that we leave the $\sigma$ integral to keep the equation as general as possible.  In order to evaluate the $\sigma$ integral one has to pick a contour on the boundary to integrate over.  The geometry of the contour determines the representation used to functionally expand $\Phi(\sigma, t,  2 \pi \emph{i} n)$ and consequently the quantum numbers that describe the CFT states.  If we take the infinite radius limit of the original BTZ coordinates to define the boundary contour we see that the metric has two independent subgroups, one that only depends on $\tilde{t}$ and one on $\tilde{\phi}$, respectively.  We can can then use separation of variables on $\Phi(\sigma, t,  2 \pi \emph{i} n)$ to write it as $\Phi(\sigma, t,  2 \pi \emph{i} n)=\sum_K a_k(\sigma) g_K(\tilde{t})h_K(\tilde{2 \pi \emph{i} n})$ and get the usual result that a BTZ black hole corresponds to two different copies of the CFT.  Any boundary contour will in general suffice and in the case of a different contour the resulting CFT states will look different depending on the representation chosen to expand $\Phi(\sigma, t,  2 \pi \emph{i} n)$.

\section{Smearing Functions}
In our previous work we claimed the interpretation of the duality in terms Green's functions and the boundary value of a field $\phi_0$ should perhaps be viewed as the relationship between a bulk function and its Fourier dual which has compact support on the boundary, with the Green's functions being the weights.  The relationship between the bulk field and the boundary operator is defined in terms of smearing functions which can be calculated from the Green's functions by taking the normal derivative. The Green's functions with the weight $\sigma$ can be read off of equation \ref{grn}, and in the outer region goes as:
\begin{equation}
(\sqrt{(\emph{l}^2 \frac{(r^2 - r_-^2))}{(r_-^2 - r_+^2)}}
   \cosh(\tilde{\phi}(t, \phi)-\tilde{\phi'}(t', 2 \pi \emph{i} n)) -
 \sqrt{(\emph{l}^2 \frac{(r^2 - r_+^2))}{(r_-^2 - r_+^2)}}
   \cosh(\tilde{t}(t, \phi)-\tilde{t}(t', 2 \pi \emph{i} n)))^{-\sigma-2}.
\end{equation}
To get the smearing functions we need to take the radial derivative of the bulk function of this.  Only the term under the square root gets differentiated and this is:
\begin{equation}
\nabla_r A(r)^{\frac{-\sigma-2}{2}} = \frac{1}{-M+\frac{r}{\emph{l}}^2+\frac{J}{4 r^2}}\frac{\partial A(r)^ {\frac{-\sigma-2}{2}}}{\partial r}=\frac{1}{-M+\frac{r}{\emph{l}}^2+\frac{J}{4 r^2}}\frac{\emph{l}^2 r (-2 - \sigma)}{(-r_-^2 + r_+^2)} (\emph{l}^2 \frac{r^2 - r_-^2}{-r_-^2 + r_+^2})^{-2-\frac{\sigma}{2}}.
\end{equation}
The only difference for the $B(r)$ terms is the replacement of $r_-$ by $r_+$ in the numerator of the final factor.  It taking limits it is helpful at this point to recall that $J$ and $M$ arise as integrals of motion and can be expressed as follows:
\begin{equation}
M=\frac{1}{\emph{l}^2}(r_+^2+r_-^2) \ \; , \; \ |J|=\frac{2}{\emph{l}} r_+r_-.
\end{equation} 
We can explore the smearing functions as we approach the two horizons and find that it blows up at each horizon.  Interestingly it is regular and vanishes at the singularity.  This is consistent with option two presented in \cite{Hamilton:2007wj} in regards to how to sensibly modify the smearing functions at finite $N$.  Unlike the example presented there no regularization scheme is needed to remove the divergence at the singularity.  Because the smearing functions are raised to the complex $\sigma$, they have an oscillating part and an exponential part, consistent with known results regarding quasi-normal modes.

\section{Black Hole Entropy}
One of the great triumphs of the AdS/CFT is its success in calculating the entropy of the BTZ black hole in the CFT (see \cite{Carlip:2005zn} for a review).  The calculation makes use of the Cardy formula which yields the density of states in a CFT.  The formula, however, depends on general properties of the CFT and not on the particulars of the individual states contributing to the entropy.  This leaves open the following questions: where do black hole micro-states lie and what exactly are those micro-states.  There are numerous answers to the latter question and the former is split between either the horizon or the boundary of the space (see \cite{Carlip:2008rk} for a review of both these questions).  Here we make an attempt to answer both these questions in light of the Fourier interpretation of the AdS/CFT.

The Fourier interpretation replaces the boundary values of fields appearing in the duality with the Fourier coefficients.  This change allows us to determine explicitly the CFT operators as well as the the states they create exactly.  It is these boundary states that compromise the micro-states of a bulk scalar field and contribute to its entropy.  By modifying the Fourier transformation on AdS we can exactly produce the CFT states corresponding to a scalar field in a black hole background.  Again, these are the states that contribute to the field's entropy, and again we see that they are states that live on the boundary.  By coupling this knowledge with the knowledge of how black holes interact with a scalar field in the bulk theory we can hope to deduce information about black holes in the CFT.  

Because black holes can interact with scalar fields, and the micro-states of a scalar field are represented as boundary states, we can conclude that the micro-states corresponding to a black hole must be states on the boundary.  Because the CFT is defined on the boundary in the Fourier interpretation, it is reasonable to assume any representation of black hole micro-states must only have support on the boundary.  Now that we determined where the micro-states are we need to ask what they are.  As stated earlier, there are a number of answers to this question.  Therefore to offer up an intuitive answer we feel we should ask the question, what must these micro-states do?  

To answer this question we look at how scalar fields interact with black holes in the bulk.  We know that a scalar field can fall into a black hole, therefore the black hole representation must be able to annihilate states of the scalar field.  We also know the black hole can emit a scalar field, as is the case in super-radiance and Hawking emission, so they must be able to create states of the scalar field as well.  Since any particle can fall in, and Hawking emission is thermal, it seems natural to suggest that rather than acting on states of a single scalar field black holes instead act on a continuous (since any mode is allowed) Fock space of states.  They should be represented in the CFT as creation/annihilation operators that raise or lower the number of particles in a given mode(s) of the Fock space.  To better understand how to represent them then we need to construct the Fock space.

The Fourier transformation tells us how to represent states of a scalar field in the CFT: they are irreducible representations of the continuous principle series of $SL(2,R)$.  Fock spaces of such states have been studied in the context of string theory, see \cite{Bars:1995cn}, \cite{Bars:1995mf} and \cite{Satoh:1997xe} for example, and we can adapt their results to our purposes.  The Fock spaces are constructed from oscillators in \cite{Bars:1995mf} as follows:
\begin{equation}
\prod_{l=1}^{\infty}{(\alpha^+_{-l})}^{a_l}\prod_{m=1}^{\infty}{(\alpha^-_{-m})}^{b_m}\prod_{n=1}^{\infty}{(s_{-n})}^{c_n}|p^-,p^+,s_0>
\end{equation}
where $a_l, b_m, c_n$ are integers and each state $|p^-,p^+,s_0>$ is a representation of the principal series of $SL(2,R)$, exactly the $\Phi(\sigma; x)$ that appear in the transform.  Comparing our notation to
theirs, we see that $s_0$ corresponds to our $\rho$.  The basis appearing in \cite{Satoh:1997xe} replaces $\rho$ with $p^2$, where $p^2 \in \mathbb{R}$.  $p^2$ 
should be interpreted as $\sqrt{\frac{2}{k-2}}\rho$
where $k$ is related to the central charge by $c=\frac{ 3 k}{k-2}$.  Recall $\rho$ is related to to the weight of the principal series representation via $\sigma=a+ \textit{i}\rho$ and corresponds to the energy of the
representation.  The $p$'s are the momenta of the string coordinates (that live in AdS) which we interpret as boundary coordinates.  The correspondence between the Fock space of string states and our CFT states is thus established.  Taking the quotient space to pass to the black hole geometry results in the quantization and analytic continuation of one the boundary coordinates (momenta).  We can now speculate on what kind of objects will fit our requirements for representing black holes in the CFT.  

We can talk about black hole entropy whether or not a scalar field is present, so a black hole should be represented by its own field.  Because it can create/annihilate states of the scalar field,  however, whatever field represents a black hole should be able to be written in terms of operators that act on the scalar Fock space.  The Virasoro generators can do exactly this.  Using results from \cite{Bars:1995mf}, we see that we can write the total generators as:
\begin{equation}
L_n= L_n^\pm +L_n^S
\end{equation}
where 
\begin{eqnarray}
L_n^\pm = \sum_m : \alpha_{-m}^-\alpha_{n+m}^+: , \\ \nonumber
L_n^S=\frac{1}{k-2}(\sum_m : s_{-m}s_{n+m}: + \textit{i} n s_n+\frac{1}{4}\delta_{n,0})
\end{eqnarray}

  It is natural then to suggest then operators representing black holes in the CFT can be represented as Virasoro Generators written in this fashion, with the Virasoro generators representing a gauge field.  The relationship between $2+1$ dimensional gravity and gauge theories is well known.  Since the generators act on the vacuum to create/annihilate states, it is reasonable to suggest that the micro-states contributing to black hole entropy are the possible external field states they can emit into.  Perhaps the true representation of a black hole will be a weighted sum (e.g. a thermal distribution) of Virasoro generators peaked around the black hole temperature.  Finding such a composite operator in terms of the representations proposed here is a subject of future work, as is understanding a stringy description of this set-up since the Fock space does after all also describe string states.
  
Our proposal for the location of black hole entropy is reminiscent of the usual black body radiation case where the states contributing to the entropy are states of the thermal photon gas: the states the black body is emitting into.  This reinforces and perhaps explains why the attempt to model the black hole as a non-interacting gas of scalar quasi-particles in the CFT in \cite{Iizuka:2002wa} was successful: according to the Fourier interpretation, this is the natural description.

\section{Conclusions}

We start with the assumptions that the Fourier Transformation on AdS defines the dual CFT and AdS black holes can be found by taking quotient spaces of AdS.  We then modify the Fourier Transformation on AdS in a manner consistent with these assumptions to find the transformation in a black hole background.  From this we read off the Green's functions which we use to calculate the smearing functions.  We find that they diverge at the two horizons and are regular at the singularity.  We also find that all three regions are mapped to the same support on the boundary, potentially paving the way for states inside the horizon to communicate with those outside the horizon via the boundary.  Our results are consistent with the literature on the subject and resolve the issue of regularity at the singularity.

The modification also yields the exact CFT micro-states of a bulk scalar field that contribute to its entropy.  They are states transforming under a modification of the principal series representation of $SL(2,R)$, with one boundary coordinate (momenta) being analytically continued and quantized.  We then borrow results from string theory to construct a Fock space of such states.  

We couple the knowledge of how black holes interact with a scalar field in the bulk theory with the new knowledge of the exact states of the external field to deduce how to represent a black hole in the CFT.  We do so by asking what a representation of the black hole must do to the micro-states of the scalar field in the CFT.  We deduce that black holes should be represented by the Virasoro generators written in terms of creation/annihilation operators that act on the Fock space of the external scalar field.  The result is that we see black hole micro-states live on the boundary and correspond to the possible modes of emission of the black hole, similar to the situation in standard black body radiation.

\section{Acknowledgments}

J. Bagger, K. Rehermann and M. Son for many useful discussions.  This work was supported in part by the US National Science Foundation, grant NSF-PHY-0401513


\begin{thebibliography}{99}


\bibitem{Banados:1992gq}
  M.~Banados, M.~Henneaux, C.~Teitelboim and J.~Zanelli,
  Phys.\ Rev.\  D {\bf 48}, 1506 (1993)
  [arXiv:gr-qc/9302012].


\bibitem{Tolfree:2008am}
  I.~M.~Tolfree,
  Phys.\ Rev.\  D {\bf 78}, 106002 (2008)
  [arXiv:0809.0485 [hep-th]].

\bibitem{Steif:1993zv}
  A.~R.~Steif,
  Phys.\ Rev.\  D {\bf 49}, 585 (1994)
  [arXiv:gr-qc/9308032].

\bibitem{Lifschytz:1993eb}
  G.~Lifschytz and M.~Ortiz,
  Phys.\ Rev.\  D {\bf 49}, 1929 (1994)
  [arXiv:gr-qc/9310008].

\bibitem{Hamilton:2006fh}
  A.~Hamilton, D.~N.~Kabat, G.~Lifschytz and D.~A.~Lowe,
  Phys.\ Rev.\  D {\bf 75}, 106001 (2007)
  [Erratum-ibid.\  D {\bf 75}, 129902 (2007)]
  [arXiv:hep-th/0612053].


\bibitem{Hamilton:2007wj}
  A.~Hamilton, D.~N.~Kabat, G.~Lifschytz and D.~A.~Lowe,
  arXiv:0710.4334 [hep-th].

\bibitem{Guijosa:2003ze}
  A.~Guijosa and D.~A.~Lowe,
  Phys.\ Rev.\  D {\bf 69}, 106008 (2004)
  [arXiv:hep-th/0312282].

\bibitem{Carlip:2005zn}
  S.~Carlip,
  Class.\ Quant.\ Grav.\  {\bf 22}, R85 (2005)
  [arXiv:gr-qc/0503022].

\bibitem{Carlip:2008rk}
  S.~Carlip,
  arXiv:0807.4192 [gr-qc].

\bibitem{Bars:1995cn}
  I.~Bars,
  arXiv:hep-th/9511187.

\bibitem{Bars:1995mf}
  I.~Bars,
  Phys.\ Rev.\  D {\bf 53}, 3308 (1996)
  [arXiv:hep-th/9503205].

\bibitem{Satoh:1997xe}
  Y.~Satoh,
  Nucl.\ Phys.\  B {\bf 513}, 213 (1998)
  [arXiv:hep-th/9705208].

\bibitem{Iizuka:2002wa}
  N.~Iizuka, D.~N.~Kabat, G.~Lifschytz and D.~A.~Lowe,
  Phys.\ Rev.\  D {\bf 67}, 124001 (2003)
  [arXiv:hep-th/0212246].

\end{thebibliography}
\end{document}